\documentclass[final,1p,times]{elsarticle} 
\usepackage{graphicx}
\usepackage{amssymb} 
\usepackage{amsthm} 
\usepackage{lineno}

\journal{Nuclear Physics A} 

\begin{document}

\begin{frontmatter} 

% Your Title - please insert
\title{Anisotropic hydrodynamics}

%% Single author (and collaboration) - please insert
%%\author{Gaius Julius Caesar (for the EMPIRE\fnref{col1} Collaboration)}
%%\fntext[col1] {A list of members of the EMPIRE Collaboration and acknowledgements can be found at the end of this issue.}
%%\address{St. Angelos Castle, Rome}

%% Multiple authors
\author[auth1,auth2]{W. Florkowski}
\author[auth3]{M. Martinez}
\author[auth1]{R. Ryblewski}
\author[auth4a,auth4b]{M. Strickland}
\address[auth1]{The H. Niewodnicza\'nski Institute of Nuclear Physics, Polish Academy of Sciences, PL-31342 Krak\'ow, Poland}
\address[auth2]{Institute of Physics, Jan Kochanowski University, PL-25406~Kielce, Poland}
\address[auth3]{Departamento de F\'isica de Part\'iculas and IGFAE, Universidade de Santiago de Compostela, E-15782 Santiago de Compostela, Galicia, Spain}
\address[auth4a]{Physics Department, Gettysburg College, Gettysburg, PA 17325 United States}
\address[auth4b]{Frankfurt Institute for Advanced Studies, Ruth-Moufang-Strasse 1, D-60438, Frankfurt am Main, Germany}

\begin{abstract} 
The recently formulated framework of anisotropic hydrodynamics is used in 3+1 dimensions to study behavior of matter created in relativistic heavy-ion collisions. The model predictions for various hadronic observables show that the effects of the initial anisotropy of pressure may be compensated by appropriate adjustment of the initial energy density. In this way, the final hadronic observables become insensitive to the early stage dynamics and the early thermalization/isotropization puzzle may be circumvented.
\end{abstract} 

\end{frontmatter} % do not change

%% linenumbers are useful for reviewing process
%\linenumbers

\section{Introduction}

Soft-hadronic observables measured in the relativistic heavy-ion experiments are well reproduced by perfect-fluid hydrodynamics or by viscous  hydrodynamics with a small viscosity to entropy ratio \cite{Chaudhuri:2006jd,Dusling:2007gi,Luzum:2008cw,Song:2007fn,Bozek:2009dw,Schenke:2010rr}. Nevertheless, the use of such approaches at the very early stages of the collisions encounters conceptual difficulties. Thermalization times shorter than a fraction of a fermi (used in the perfect-fluid approaches) cannot be explained within microscopic models of the collisions. On the other hand, viscous hydrodynamics is based on an implicit assumption that one can make an expansion around an  isotropic background. If the shear correction is large, a new framework incorporating  large momentum-space anisotropies into the leading order of the approximation may be useful. Such a new approach has been introduced in \cite{Florkowski:2010cf,Martinez:2010sc,Ryblewski:2010bs,Martinez:2010sd,Ryblewski:2011aq,Martinez:2012tu,Ryblewski:2012rr} and we refer to it below as to the {\it anisotropic hydrodynamics}.

\section{Anisotropic hydrodynamics}

Anisotropic hydrodynamics is based on the equations
\begin{eqnarray}
\partial_\mu T^{\mu \nu} &=& 0, \label{enmomcon} \\
\partial_\mu \sigma^{\mu} &=& \Sigma, \label{engrow}
\end{eqnarray}
which express the energy-momentum conservation and entropy production laws. The energy-momentum tensor $T^{\mu \nu}$ has the structure
\begin{eqnarray}
T^{\mu \nu} = \left( \varepsilon  + P_{\perp}\right) U^{\mu}U^{\nu} - P_{\perp} \, g^{\mu\nu} - (P_{\perp} - P_{\parallel}) V^{\mu}V^{\nu}, 
\label{Taniso}
\end{eqnarray}
where $P_{\parallel}$ is the longitudinal pressure and $P_{\perp}$ is the transverse pressure. In the limit $P_{\parallel}=P_{\perp}=P$, Eq.~(\ref{Taniso}) reproduces the energy-momentum tensor of the perfect fluid. Similarly, the entropy production law (\ref{engrow}) is reduced to the entropy conservation law, if we take $\Sigma=0$. The four-vector $U^{\mu}$ describes the flow of matter, while $V^{\mu}$ defines the beam ($z$) axis. In the general case, we use the parameterizations $U^\mu = (u_0 \cosh \vartheta, u_x, u_y, u_0 \sinh \vartheta)$ and $V^\mu = (	 \sinh \vartheta, 0, 0,  \cosh \vartheta)$, where $u_x$ and $u_y$ are the transverse components of the four-velocity field and $\vartheta$ is the longitudinal fluid rapidity. The entropy flux $\sigma^{\mu}$ equals $\sigma \, U^\mu$, where $\sigma$ is the non-equilibrium entropy density.

One can show \cite{Florkowski:2010cf} that instead of $P_{\parallel}$ and $P_{\perp}$ it is more convenient to use the entropy density $\sigma$ and the {\it anisotropy parameter} $x$ as two independent variables (one may use the approximation $P_{\parallel}/P_{\perp} \approx x^{-3/4}$). Similarly to standard hydrodynamics with vanishing baryon chemical potential, the energy density $\varepsilon$, the entropy density  $\sigma$, and the anisotropy parameter $x$ are related through the \textit{generalized} equation of state  \cite{Ryblewski:2010bs}
\begin{eqnarray}
\varepsilon (x,\sigma)&=&  \varepsilon_{\rm qgp}(\sigma) r(x), \label{epsilon2b}  \\ \nonumber 
P_\perp (x,\sigma)&=&  P_{\rm qgp}(\sigma) \left[r(x) + 3 x r^\prime(x) \right], \label{PT2b}   \\ \nonumber 
P_\parallel (x,\sigma)&=&  P_{\rm qgp}(\sigma) \left[r(x) - 6 x r^\prime(x) \right]. \label{PL2b} 
\end{eqnarray}
where the functions $\varepsilon_{\rm qgp}$ and  $P_{\rm qgp}$  define the realistic QCD equation of state constructed in Ref. \cite{Chojnacki:2007jc}. The function $r(x)$ characterizes  properties of the fluid which exhibits the pressure anisotropy $x$  \cite{Florkowski:2010cf}
\begin{equation}
r(x) = \frac{ x^{-\frac{1}{3}}}{2} \left[ 1 + \frac{x \arctan\sqrt{x-1}}{\sqrt{x-1}}\right].
\label{RB}
\end{equation}
In the isotropic case $x = 1$, $r(1)=1$, $r^\prime(1)=0$, and Eq.~(\ref{epsilon2b}) is reduced to the standard equation of state used in \cite{Chojnacki:2007jc}.

The function $\Sigma$ in Eq.~(\ref{engrow}) defines the entropy source. We use the form proposed in \cite{Florkowski:2010cf}
\begin{equation}
\Sigma(\sigma,x) = \frac{(1-\sqrt{x})^{2}}{\sqrt{x}}\frac{\sigma}{\tau_{\rm eq}},
\label{entropys}
\end{equation}
where the time-scale parameter $\tau_{\rm eq}$ controls the rate of equilibration. In this work we use the constant value $\tau_{\rm eq}$ = 1 fm. In Refs. \cite{Martinez:2010sc,Martinez:2010sd,Martinez:2012tu} the medium dependent $\tau_{\rm eq}$ was used, which was inversely proportional to the typical transverse momentum scale in the system. If a constant value of $\tau_{\rm eq}$ is used, the system approaches the perfect fluid behavior for $\tau \gg \tau_{\rm eq}$. 

In the limit of small anisotropy Eq.~(\ref{entropys}) is consistent with the quadratic form of the entropy production in the Israel-Stewart theory \cite{Martinez:2010sc}. Far from equilibrium, hints for the form of $\Sigma$ are lacking, although we may expect some suggestions from the AdS/CFT correspondence \cite{Heller:2011ju}. Thus, for large anisotropies the formula (\ref{entropys}) should be treated as an assumption defining the dynamics of the system. 

\section{Initial conditions and freeze-out}

In the general 3+1 dimensional case we have to solve Eqs. (\ref{enmomcon}) and (\ref{engrow}) for $\sigma$, $x$, $u_x$, $u_y$, and $\vartheta$, which depend on $\tau,{\bf x}_\perp=(x,y)$, and $\eta$ ($\tau$ is the proper time and $\eta$ is the space-time rapidity). We fix the initial starting time to $\tau_{\rm 0} =0.25$ fm. Similarly to other hydrodynamic calculations, we assume that there is no initial transverse flow, $u_x(\tau_{\rm 0},{\bf x}_\perp,\eta) = u_y(\tau_0,{\bf x}_\perp,\eta) = 0$ and that the initial longitudinal rapidity of the fluid is equal to space-time rapidity, $\vartheta(\tau_0,{\bf x}_\perp,\eta) = \eta$. In this text we present the results for two scenarios: i) the initial source is strongly {\it oblate} in momentum space, $x(\tau_0,{\bf x}_\perp,\eta) =100$, and ii) the source is {\it prolate} in momentum space, $x(\tau_0,{\bf x}_\perp,\eta) = 0.032$; the latter value is chosen since $r(100) = r(0.032)$. The initial entropy density profile has the form
\begin{equation}
 \sigma_0(\eta,{\bf x}_\perp) = \sigma(\tau_0,\eta,{\bf x}_\perp) = \varepsilon_{\rm gqp}^{-1} 
\left[ \varepsilon_{\rm i} \, \tilde{\rho}(b,\eta,{\bf x}_\perp) \right],
\label{sig2}
\end{equation}
where $b$ is the impact parameter, and $\tilde{\rho}(b,\eta,{\bf x}_\perp)$ is the normalized density of sources, $\tilde{\rho}(b,\eta,{\bf x}_\perp) = \rho(b,\eta,{\bf x}_\perp)/\rho(0,0,0)$, for details see \cite{Ryblewski:2012rr}. The quantity $\varepsilon_{\rm i}$ is the initial energy density at the center of the system created in the most central collisions. Its value is fixed by the measured multiplicity, separately for two different physical scenarios considered in this paper. We use $\varepsilon_{\rm i}$ = 48.8 GeV/fm$^3$ and 80.1 GeV/fm$^3$ for $x_0 = 100$ and $x_0 = 0.032$, respectively.

\begin{figure}[t]
\begin{center}
\includegraphics[angle=0,width=0.5\textwidth]{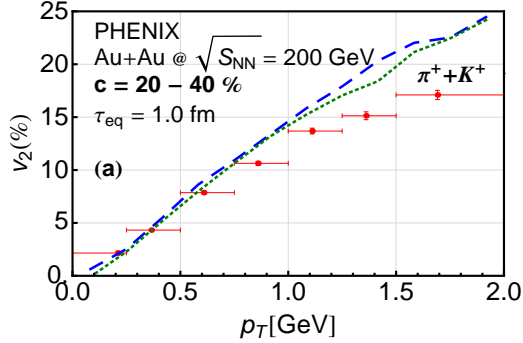}
\caption{\small (Color online) Transverse-momentum dependence of the elliptic flow coefficient $v_2$ of $\pi^{+}+K^{+}$ calculated for $c=20-40$\% ($b=7.84$ fm) at midrapidity and for $\tau_{\rm eq}= 1.0$ fm/c; $x_0=100$ (dashed blue lines) and $x_0=0.032$ (dotted green lines). The results are compared to the PHENIX Collaboration data (red dots) \cite{Adler:2003kt}. 
}
\end{center}
\label{fig1}
\end{figure}

\begin{figure}[t]
\begin{center}
\includegraphics[angle=0,width=0.5\textwidth]{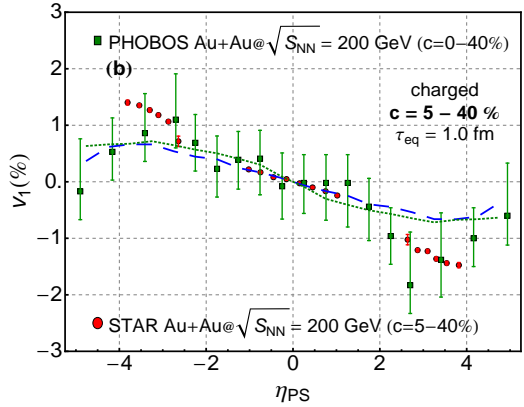}
\caption{\small (Color online) Pseudorapidity dependence of the directed flow of charged particles for the centrality bin $c=5-40$\% and $\tau_{\rm eq}=1.0$ fm/c;  $x_0=100$ (dashed blue lines) and $x_0=0.032$ (dotted green lines). The results are compared to the experimental data from STAR (red dots) \cite{Abelev:2008jga} and PHOBOS (green squares) \cite{Back:2005pc}. 
}
\end{center}
\label{fig2}
\end{figure}

The evolution is determined by the hydrodynamic equations until the entropy density drops to $\sigma_{\rm f} = 1.79$ fm$^{-3}$, which for $x=1$ corresponds to the temperature $T_{\rm f} = 150$ MeV. According to the single-freeze-out scenario, at this moment the abundances and momenta of particles are expected to be fixed. The processes of particle production and decays of unstable resonances are described by using  {\tt THERMINATOR} \cite{Kisiel:2005hn,Chojnacki:2011hb}, which applies the Cooper-Frye formalism to generate hadrons on the freeze-out hypersurface.

\section{Results}

 The model results describing the pseudorapidity distributions, transverse-momentum spectra, the elliptic and directed flow coefficients, and the HBT radii have been obtained with different initial anisotropies of pressure; $x_0=100$ (dashed blue lines) and $x_0=0.032$ (dotted green lines). In all of the considered cases we find good agreement between the model results and the data. Moreover, we find that the results obtained with different initial anisotropies are practically the same. This is so because we have adjusted the initial energy density separately for two different values of $x_0$. A larger (smaller) initial energy density is used for the initially prolate (oblate) system. Our results describing the elliptic and directed flow are shown and compared to the RHIC data (Au+Au collisions at the highest beam energy $\sqrt{s_{\rm NN}}$ = 200 GeV) in Figs.~1 and 2. 

Our results indicate that the final hadronic observables are not sensitive to the early anisotropy of pressure. The flows are built up during the whole time evolution of the system, hence the relatively short early anisotropic stage does not influence the results. In our opinion, the insensitivity of the hadronic observables helps us to circumvent the early thermalization/isotropization puzzle.

\medskip 

{\bf Acknowledgements:} This work was supported by the Polish Ministry of Science and Higher Education under Grant No.~N N202 263438 and the United States National Science Foundation under Grant No. PHY-1068765.

\end{document}